\documentclass{llncs}

\usepackage{amsmath}
\usepackage{amssymb}
\usepackage{fancyvrb}
\usepackage{oz}
\usepackage{mathtools}
\usepackage{proof}
\usepackage{array}
\usepackage{wrapfig}
\usepackage{color}
\usepackage[space]{cite}

\newcommand{\codesize}{\footnotesize}

\newcommand{\seqslice}{\mathbin{\downharpoonright}}
\newcommand{\seqcomp}{}
\newcommand{\sq}[1]{\langle#1\rangle}

\newcommand{\stackarrow}[1]{\xrightarrow{\scriptstyle #1}}

\newcommand{\porel}{\stackarrow{p}}

\newcommand{\rsrel}{\stackarrow{RS}}

\newcommand{\deprel}{\hookrightarrow}

\newcommand{\thract}{thr\_act}

\newcommand{\fnup}[3]{#1\oplus\{#2\mapsto#3\}}

\newcommand{\Proof}{{\bf Proof.~}}

\mathchardef\hyph="2D

\title{Linearizability and Causality}

\author{Simon Doherty \and John Derrick}
\institute{Department of Computing, University of Sheffield, Sheffield, UK}

\begin{document}

\maketitle

\begin{abstract}
Most work on the verification of concurrent objects for shared memory assumes sequential 
consistency, but most multicore processors support only {\em weak memory models}
that do not provide sequential consistency. Furthermore, most verification efforts
focus on the {\em linearizability} of concurrent objects, but there are existing implementations
optimized to run on weak memory models that are not linearizable.

In this paper, we address these problems by introducing {\em causal linearizability}, a correctness
condition for concurrent objects running on weak memory models. Like linearizability itself,
causal linearizability enables concurrent objects to be composed, under weak constraints on
the client's behaviour. We specify these constraints by introducing
a notion of {\em operation-race freedom}, where programs that satisfy this property
are guaranteed to behave as if their shared objects were in fact linearizable.

We apply these ideas to objects from the Linux kernel, optimized to run on TSO, the
memory model of the x86 processor family.
\end{abstract}

\section{Introduction}

The past decade has seen a great deal of interest in the verification of highly optimized, shared-memory concurrent
objects. This interest is partly motivated by the increasing importance of multicore systems.
Much of this verification work has assumed that these concurrent implementations run on the sequentially
consistent memory model. However, contemporary multicore architectures do not implement this strong model. Rather, they
implement {\em weak memory models}, which allow reorderings of memory operations, relative to what would be legal under
sequential consistency. Examples of such models include TSO (implemented on the x86) \cite{owens-x86-tso},
POWER and ARM \cite{Alglave:2009:SPA:1481839.1481842}.
These models create significant challenges for verifying that an implementation satisfies a particular
correctness condition \cite{derrick-TSO-lin-2014}.

Furthermore it is not always clear {\em what} 
correctness conditions are appropriate for an implementation running on a weak memory model.
Specifically, the standard correctness condition for concurrent objects is {\em linearizabilty} \cite{linearizability-toplas}. However, as
described in Section \ref{sect:non-lin-objects}, there are implementations of concurrent objects optimized to run on weak
memory models that are not linearizable. Nevertheless, these implementations are used in important contexts, including the
Linux kernel. This is possible because when these objects are used in a stereotypical fashion, their nonlinearizable behaviours
are not observable to their clients. 

Our goal in this paper is to define a correctness condition appropriate for these nonlinearizable objects.
We introduce a correctness condition called {\em causal linearizablilty}. Roughly speaking, an object is
causally linearizable if all its executions can be transformed into linearizable executions, in a way that is not
observable to any thread.
As we shall see, causal linearizability is stronger than sequential consistency, and therefore programmers can reason
about causally linearizable systems using established intuitions and verification techniques.
Furthermore, unlike some competing proposals, causal linearizability places no
constraints on the algorithmic techniques used in the implementation of concurrent objects.

Causal linearizability enables concurrent objects to be composed, under certain constraints on
the client's behaviour. We specify these constraints by introducing
a notion of {\em operation-race freedom}, where programs that satisfy this property
are guaranteed to behave as if their shared objects were linearizable.



In the remainder of the introduction we motivate our work by describing a nonlinearizable data structure designed for
a particular weak memory model (in this case, TSO).
The structure of the rest of the paper is as follows.
Section \ref{sect:contrib} outlines our contribution, and compares it to related work.
Section \ref{sec:threadshistobjects} defines the formal framework and notation, and  Section \ref{sec:independence} defines independence and causal ordering, which are key concepts in our definition of causal linearizability.
Section \ref{sect:clients} defines causal linearizability itself.
Section \ref{sect:flush-mem} then defines operation-race freedom and outlines a proof method for proving causal linearizability.
Section \ref{sect:tso-causal} applies our ideas to the TSO memory model.
Section \ref{sec:conclusions} concludes.



\subsection{Motivation - Nonlinearizable Objects on TSO}
\label{sect:non-lin-objects}

The Total Store Order (TSO) memory model optimizes write operations by first {\em buffering} a write to a local {\em write buffer}, and later
{\em flushing} the write to shared memory. The effect of the write is immediately visible to the core that issues it, but is only visible
to other cores after the write has been flushed. The x86 instruction set provides primitives for ensuring that the effect of a write is
visible to other threads on other cores. The {\em barrier} operation flushes all writes of the executing core
that have not previously been flushed. In addition, memory operations that both read and modify shared memory
may be {\em locked}. A locked operation appears to execute atomically, and the locking mechanism causes the executing core's
write buffer to be emptied both before and after the execution of the locked operation. We formalize this memory model in Section 
\ref{sect:tso-causal}.

Locked operations and barriers are typically costly, relative to simple reads and writes.
For this reason, optimized datastructures often avoid such synchronization primitives
where possible. Here we describe a simple example of such an algorithm: a {\em spinlock} algorithm for x86 processors
that is adapted from an implementation in the Linux kernel. Figure \ref{fig:spinlock:code} presents
pseudocode for the algorithm, which uses a simple boolean flag (\verb|F| below) to record whether
the lock is currently held by some thread. A thread acquires the lock using the \verb|try_acquire| procedure, which fails if the lock
is currently held (an unconditional \verb|acquire| procedure can be implemented by repeatedly invoking \verb|try_acquire| until successful).
The \verb|try_acquire| procedure uses a locked operation to atomically determine whether the lock is held and to set the flag to true.
(This operation is called an atomic {\em test-and-set}). Note that this operation has no effect if the flag is already true.
Thus, if the lock is not held, then \verb|try_acquire| successfully acquires the lock and returns \verb|true|.
Otherwise, the acquisition attempt fails, and \verb|try_acquire| returns \verb|false|.
The optimised \verb|release| operation simply sets the flag to \verb|false|, without using any locked or barrier operations.
\begin{figure}[t]
\begin{sidebyside}[2]
{\codesize
\begin{verbatim}
bool F = false;

void release() {
R1  F := false;
}
\end{verbatim}
}
\nextside
{\codesize
\begin{verbatim}
bool try_acquire() {
T1  locked {
T2      held := F;
T3      F := true;
T4   }
T5  return !held;
}
  \end{verbatim}
}
\end{sidebyside}
\vspace{-12mm}
\caption{Nonlinearizable spinlock implementation}
\label{fig:spinlock:code}
\end{figure}

The spinlock implementation is {\em not} linearizable.
Intuitively, linearizability requires that
each operation on the lock appears to take effect at some point between its
invocation and response. To see how this fails, consider
the execution in Figure \ref{fig:spinlock-non-lin}.
\begin{wrapfigure}{L}{0.5\textwidth}
\centering 
{\codesize\tt
\vspace{-1.6em}
\begin{tabular}{l|l}
\multicolumn{1}{c}{T1} & \multicolumn{1}{c}{T2}\\
\hline
L.try\_acquire() & \\
returns true & \\
L.release() & \\
return& \\
& L.try\_acquire()\\
& returns false
\end{tabular}
}
\caption{Nonlinearizable spinlock history.}
\label{fig:spinlock-non-lin}
\end{wrapfigure}
In this example, two threads, $t_1$ and $t_2$ attempt to acquire the lock \verb|L|, using the \verb|try_acquire| operation.
The first acquisition attempt (of $t_1$) succeeds, because the lock is free; $t_1$ then releases the lock (presumably after
accessing some shared state protected by the lock), but the write that changes the lock's state is not yet flushed to shared
memory. Now $t_2$'s lock acquisition fails, despite being invoked after the end of $t_1$'s \verb|release| operation.
This is because, in the example, the releasing write is not flushed until after the completion of $t_2$'s \verb|try_acquire|.
Thus, $t_2$'s \verb|try_acquire| appears to take effect between $t_1$'s acquisition and release operations. Linearizability
requires that the \verb|try_acquire| appear to take effect {\em after} $t_1$'s release.

Despite the fact that it is not linearizable, there are important circumstances in which this spinlock implementation
can be correctly used. Indeed, a spinlock essentially identical to this has been used extensively in the Linux kernel.
Fundamentally, the goal of this paper is to investigate and formalize conditions under which objects like this spinlock
may be used safely on weak memory models.

\section{Our Contribution}
\label{sect:contrib}

In this paper, we describe a weakening of linearizability which we call {\em causal linearizability}.
Note that in the execution of Figure \ref{fig:spinlock-non-lin}, no thread can observe that the invocation of thread $t_2$'s
\verb|try_acquire| occurred after the response of $t_1$'s release, and therefore no thread can observe the failure
of linearizability. Causal linearizability allows operations to be linearized {\em out of order}, in cases where no thread
can observe this reordering.

However, the lock \verb|L| is the only object in execution in Figure \ref{fig:spinlock-non-lin}. In general, clients of the lock
may have other means of communicating, apart from operations on \verb|L|. Therefore, under some circumstances, clients may observe that
the execution is not linearizable. Our second contribution is to define a condition called {\em operation-race freedom} (ORF)
on the {\em clients} of nonlinearizable objects like the TSO spinlock, such that clients satisfying ORF cannot observe
a failure of linearizability. 
ORF is inspired by {\em data-race freedom} (DRF) approaches, which we discuss below. However,
unlike DRF, ORF is defined in terms of high-level invocations and responses, rather than low-level reads and writes.

Finally, we provide a proof method for verifying causal linearizability. We define a correctness condition called
{\em response-synchronized linearizability} (RS-linearizability). In the context of TSO, proving that a data structure satisfies
RS-linearizability amounts to proving that the structure is linearizable in all executions where all buffered writes
are flushed by the time any operation completes. This means that we can prove RL linearizability
using essentially standard techniques used for proving linearizability. In Secton \ref{sect:flush-mem}, we show that
any set of RL linearizable objects is causally linearizable, when the objects' clients satisfy ORF.

\subsubsection{Related Work}

One way to address the issues raised by weak memory models is based on the observation that locks and other
synchronization primitives are typically used in certain stereotypical ways. For example, a lock is never
released unless it has been first acquired;
and the shared state that a lock protects is not normally accessed without holding the lock. As shown in \cite{triang-race-freedom},
these circumstances mean that the spinlock's nonlinearizable behaviour can never be observed by any participating thread.

The analysis given in \cite{triang-race-freedom} belongs to a class of approaches that define conditions under which a program
running on a weak memory model will behave as if it were running on a sequentially-consistent memory.
These conditions are often phrased in terms  {\em data-races}: a data-race is a pair of operations executed by different
threads, one of which is a write, such that the two operations can be adjacent in some execution of the (multithreaded) program.
Data-race free (DRF) programs are those whose executions never contain data races, and the executions of DRF programs are always
sequentially consistent.

Data-race free algorithms that are linearizable on the sequentially-consistent model will
appear to be linearizable on the appropriate weak memory model.
Thus the problem of verifying a DRF algorithm on weak-memory is reduced to that of verifying it
under the standard assumption of sequential consistency. However, DRF-based approaches have the drawback that algorithms
that are not DRF cannot be verified. Our approach does not suffer from this limitation: implementations are free to use any algorithmic
techniques, regardless of DRF. Furthermore, the ORF property only constrains the ordering of high-level invocations and responses,
rather than low level reads and writes.

\cite{tso-lib-correct-2012, gotsman2012show} define correctness conditions for TSO by weakening linearizability.
\cite{tso-lib-correct-2012} introduces abstract specifications that manipulate TSO-style write-buffers such that the abstract
effect of an operation can be delayed until after the operation's response. \cite{gotsman2012show} proposes adding nondeterminism to 
concurrent objects' specifications to account for possible delay in the effect of an operation becoming visible.
Neither work systematically addresses how to reason about the behaviour of clients that use these weakened abstract objects.
In our work, the abstract specifications underlying linearizability are unchanged, and programs satsifying the ORF constraint
are guaranteed to behave as if their shared objects were linearizable.

RL linearizability is a generalisation of {\em TSO-linearizability}, described in \cite{derrick-TSO-lin-2014}.
That work shows that TSO-linearizability can be verified using more-or-less standard techniques for proving
linearizability. However, \cite{derrick-TSO-lin-2014} does not address how to reason about the behaviour of clients that use
TSO-linearizable objects, as we do with the ORF constraint.

\section{Modelling Threads, Histories and Objects}
\label{sec:threadshistobjects}

As is standard, we assume a set of invocations $I$ and responses $R$, which are used to represent operations on a set
$X$ of {\em objects}.
The invocations and responses of an object define its interactons with the external environment so we define
$Ext = I\cup R$ to be the set of {\em external actions}.
We denote by $obj(a)$ the object associated with $a\in Ext$.
We also assume a set of {\em memory actions}
$Mem$, which typically includes reads, writes and other standard actions on shared-memory. Operational definitions of weak memory
models typically involve {\em hidden actions} that are used to model the memory system's propagation of information between
threads, so we assume a set $Hidden\subseteq Mem$ (for example, in TSO, the hidden actions are the flushes). 
Let $Act = Ext\cup Mem$ be the set of {\em actions}.

In our model, each action is executed either on behalf of a thread (e.g., invocations, or read operations),
or on behalf of some memory system (these are the hidden actions).
To represent this, we assume a set of threads $T$, a function $thr: Act\to T\cup\{\bot\}= T_{\bot}$,
such that $thr(a)=\bot$ iff $a\in Hidden$.

Executions are modelled as {\em histories}, which are sequences of actions. 
We denote by $g\seqcomp h$ the concatenation of two histories $g$ and $h$.
When $h$ is a history and $A$ a set of actions, we denote by $h\seqslice A$ the sequence
of actions $a\in A$ occurring in $h$.
For a history $h$, the {\em thread history of $t \in T$}, denoted $h\seqslice t$, is $h\seqslice \{a : thr(a)=t\}$.
Two histories $h$ and $h'$ are {\em thread equivalent} if $h\seqslice t = h'\seqslice t$, for all threads $t\in T$. (Note that two histories
may be thread equivalent while having different hidden actions.)

For example, the behaviour presented in Figure \ref{fig:spinlock-non-lin} can be represented as the following history.
\begin{multline}
\label{ex:hist1}
\hspace{-0.1cm} L.try\_acq_{t_1}, locked_{t_1}(TAS, F, false), resp_{t_1}(L, true), L.release_{t_1}, write_{t_1}(F, false),\\
\hspace{-0.1cm} resp_{t_1}(L), L.try\_acq_{t_2}, locked_{t_2}(TAS, F, true), resp_{t_2}(L, false), flush(F, false)
\end{multline}
Let $a=L.try\_acq_{t_1}$. Then $a$ is an invocation of the \verb|try_acquire| operation, $thr(a) = t_1$ and $obj(a) = L$.
The action $resp_{t_1}(L, true)$ is a response from object $L$, of the thread $t_1$, returning the value $true$.
$locked_{t_1}(TAS, F, false)$ is a locked invocation of the test-and-set operation on the location $F$, again by thread $t_1$ that returns the
value $false$. $flush(F, false)$ is a flush action of the memory subsystem, that sets the value of $F$ to false in the shared store.
This history is thread equivalent to the following:
\begin{multline}
\label{ex:hist2}
L.try\_acq_{t_1}, locked_{t_1}(TAS, F, false), resp_{t_1}(L, true),\\
L.try\_acq_{t_2}, locked_{t_2}(TAS, F, true), resp_{t_2}(L, false),\\
L.release_{t_1}, write_{t_1}(F, false), flush(F, false), resp_{t_1}(L)
\end{multline}

A history is well-formed if for all $t \in T$,
$h \seqslice t \seqslice Ext$ is an alternating sequence of invocations
and responses, beginning with an invocation. Note that well-formedness only constrains 
invocations and responses. Memory operations may be freely interleaved with the external actions.
From now on, we assume that all histories are well-formed.
A history is {\em complete} if every thread history is empty or ends in a response.

An {\em object system} is a prefix-closed set of well-formed histories. A {\em sequential object system} is
an object system where every invocation is followed immediately by a response, in every history.
If $O$ is an object system then $acts(O)$ is the set of actions appearing in any history of $O$.

We wish to reason about orders on the actions appearing in histories. In general, each action may appear several times in a history.
Strictly speaking, to define appropriate orders on the actions, we would need to tag actions with some identifying information, to obtain an {\em event} which is guaranteed to be unique in the history. However, for the sake of simplicity, we assume that each action
only appears at most once in each history. For example, each thread may only execute at most one write for each location-value pair.
This restriction can be lifted straightforwardly, at the cost of some notational complexity.
\!\footnote{The full version of the paper, which can be found at TODO, presents a model of histories in which events are unique.}

Given a history $h$, the {\em real-time order} of $h$, denoted $\to_h$ is the strict total order on actions such that $a\to_h b$ if
$a$ occurs before $b$ in $h$. The {\em program order}, denoted $\porel_h$, is the strict partial order
on the actions of $h$ such that $a\porel_h b$ if $thr(a)=thr(b)$ and $a \to_h b$.
For example, in History \ref{ex:hist1} above, $L.release_{t_1}\porel_h write_{t_1}(F, false)$ and $write_{t_1}(F, false)\to_h flush(F, false)$.


\section{Independence and Causal Ordering}
\label{sec:independence}



In this section, we develop a notion of causal ordering.
Roughly speaking, an action $a$ is causally prior to an action $b$ in a history $h$ if $a\to_h b$
and some thread can observe that $a$ and $b$ occurred in that order. Therefore, we can safely reorder events that are not
causally ordered. Causal order itself is expressed in terms of an independence relation between actions, which we now define.
The notion of independence, and the idea of using independence to construct a causal order has a long history.
See \cite{abstract-for-conc-obj-2010} for a discussion in a related context.

Given an object system $S$, two actions $a$ and $b$ are {\em $S$-independent} if $thr(a)\neq thr(b)$ and for all
histories $g$ and $h$,
\begin{align}
g\seqcomp \sq{a, b}\seqcomp h \in S \iff g\seqcomp \sq{b, a} \seqcomp h \in S
\end{align}
(Here, $\sq{a, b}$ is the sequence of length two containing $a$ and then $b$.)
According to this definition, TSO flushes are independent iff they are to distinct locations.
Again in TSO, read and write actions in different threads are always independent, but two actions of the same thread never are.
(Inter-thread communication only occurs during flush or locked actions.)

We define the causal order over a history in terms of this independence relation.
We say that $h$ is {\em $S$-causally equivalent} to $h'$ if $h'$ is obtained from $h$ by zero or more
transpositions of adjacent, $S$-independent actions. Note that causal equivalence is an equivalence relation.
Actions $a$ and $b$ are $S$-causally ordered in $h$, denoted $a\deprel_h^S b$ if for all causally equivalent histories $h'$,
$a \to_{h'} b$. This is a transitive and acyclic relation, and therefore $\deprel_h^S$ is a strict partial order.

For example, because the release operation does not contain any locked actions, Histories \ref{ex:hist1} and \ref{ex:hist2} on
page \pageref{ex:hist2} are causally equivalent. On the other hand, the actions $locked_{t_2}(TAS, F, true)$ and $flush(F, false)$
are not independent, and therefore $locked_{t_2}(TAS, F, true) \deprel_h flush(F, false)$.

Note that independence, causal equivalence, and causal order are all defined relative to a specific object system.
However, we often elide the object system parameter when it is obvious from context.

One key idea of this work is that a history is ``correct'' if it can be transformed into a linearizable history in a
way that is not observable to any thread. The following lemma is our main tool for effecting this transformation.
It says that a history can be reordered to be consistent with any partial order that contains the
history's causal ordering. The thrust of our compositionality condition, presented in Section \ref{sect:flush-mem}, is to provide sufficent
conditions for the existence of a strict partial order satisfying the hypotheses of this lemma.
\!\footnote{For reasons of space, this paper does not contain proofs of Lemma \ref{lem:causal-reordering} or the other results
presented in this paper. Proofs can be found in the full version of the paper at TODO.}

\begin{lemma}
\label{lem:causal-reordering}
Let $S$ be an object system, let $h \in S$ be a history, and let $<$ be a strict partial order on the events
of $h$ such that $\deprel_h^S\subseteq <$. Then 
there exists an $h'$ causally equivalent to $h$ such that for all events $a,b$ in $h$ (equivalently in $h'$)
$a < b$ implies $a\to_{h'} b$. \qed
\end{lemma}

We are now in a position to formally define causal linearizability. 
Essentially, an object system is causally linearizable
if all its histories have causally equivalent linearizable histories.
The key idea behind linearizablity is that each operation should appear to take effect atomically, at some point between the
operation's invocation an response.
See \cite{linearizability-toplas} or \cite{mech-verif-proof-2011} for a formal definition.
\begin{definition}[Causal Linearizability]
\label{def:causal-lin}
An object system $S$ is {\em causally linearizable} to a sequential object system $T$ if for all $h\in S$,
$h$ is $S$-causally equivalent to some history $h'$ such that $h'\seqslice acts(T)\cap Ext$ is linearizable to $T$.
\end{definition}
Note that causal linearizability is defined in terms of histories that contain both
external and internal actions. Typically linearizability and related correctness conditions are defined purely
in terms of external actions. Here, we preserve the internal actions of the object, because those internal actions
carry the causal order.

\section{Observational Refinement and Causal Linearizability}
\label{sect:clients}

In this section, we introduce a notion of {\em client} and a notion of composition of a
client with an object system (Definition \ref{def:compo}).
We then define a notion of {\em observational refinement} for object systems. One object system $S$ observationally refines
another object system $T$ for a client $C$ if the external behaviour of $C$ composed with $S$ is included in the external
behaviour of $C$ composed with $T$.
These notions have a twofold purpose. First, they provide a framework in which to show that causal linearizability is a reasonable correctness
condition: the composition of a client with a causally linearizable object system has only the behaviours of the client
composed with a corresponding linearizable object system (Theorem \ref{theorem:causal-lin-sound}). Second, 
these notions allow us to specify a constraint on the behaviour of a client, such that the client can safely use
a composition of nonlinearizable objects.

A {\em client} is a prefix-closed set of histories, where each history contains
only one thread, and all actions are thread actions (so that the client contains no hidden actions).
Each client history represents a possible interaction of a client thread with an object system.
While each client history contains only one thread, the client itself may contain histories of several threads.
For example, consider the histories that might be generated by a thread $t_1$ repeatedly executing
spinlock's \verb|try_acquire| operation
(Figure \ref{fig:spinlock:code}) until the lock is
successfully acquired.
The set of histories generated in this way for every thread is a client. One such history is
$L.try\_acq_{t_1}, locked_{t_1}(TAS, F, false), resp_{t_1}(L, true)$,
where $t_1$ successfully acquires the lock on the first attempt.
A history where the thread acquires the lock after two attempts is
\begin{multline}
L.try\_acq_{t_1}, locked_{t_1}(TAS, F, true), resp_{t_1}(L, false), \\
L.try\_acq_{t_1}, locked_{t_1}(TAS, F, false), resp_{t_1}(L, true)
\end{multline}
Thus, the client histories contain the memory operations determined by the implementations of the shared objects.

The composition of an object system $O$ and client
program $C$, denoted $C[O]$ is the object system defined as follows:
\begin{align}
\label{def:compo}
C[O] = \{h : h\seqslice acts(O)\in O \wedge  \forall t\in T.~h\seqslice t\in C\}
\end{align}
So for all $h\in C[O]$, $h$ is an interleaving of actions of the threads in $C$, and
every thread history of $h$ is allowed by both the object system and the client program.

We need a notion of observational refinement relative to a given client.
\begin{definition}[Observational Refinement]
An object system $S$ {\em observationaly refines} an object system $T$ for a client $C$ if for every $h \in C[S]$,
there exists some $h' \in C[T]$ where $h\seqslice Ext$ and $h'\seqslice Ext$ are thread equivalent.
\end{definition}

The following theorem shows that causal linearizability is sound with respect to observational refinement.
\begin{theorem}[Causal Linearizability Implies Observational Refinement]
\label{theorem:causal-lin-sound}
Let $T$ be a sequential object system, and let $T'$ be its set of linearizable histories. Let $S$ be an object system
such that $acts(T) \cap Ext = acts(S) \cap Ext$. If $C[S]$ is causally linearizable to $T$, then $S$ observationally refines
$T'$ for $C$.
\end{theorem}


\section{Flush-based Memory and Operation-race Freedom}
\label{sect:flush-mem}

Causal linearizability is a general correctness condition, potentially applicable in a range of contexts.
Our goal is to apply it to objects running on weak memory models. To this end, we formally
define a notion of {\em flush-based memory}. Flush-based memory is a generalisation of TSO and some other memory
models, including {\em partial store order} \cite{weak-memory-a-tutorial}. This section develops a
proof technique for causal linearizability of an object system running on flush-based memory, and hence for observational refinement.

Our proof technique can be encapsulated in the following formula:
{\em Operation-race freedom $+$ Response-synchronized linearizabilty $\implies$ Causal linearizability}.
Response-synchronized linearizability, a weakening of linearizability, is a correctness property
specialised for flush-based memory, and is adapted from {\em TSO linearizability} studied in \cite{derrick-TSO-lin-2014}.
That work presents techniques for verifying TSO linearizability and proofs that spinlock and seqlock are TSO linearizable.
Theorem \ref{theorem:composition} below shows that a multi-object system composed of response-synchronized
linearizable objects is causally linearizable, under a constraint on the multi-object system's clients. This constraint
is called {\em operation-race freedom}, given in Definition \ref{def:operation-race}.

A {\em flush-based memory} is an object system whose histories do not contain invocations or responses (so its only
actions are memory actions), together with a {\em thread-action} function $\thract_h: Hidden \to Act$,
for each history $h$ in the memory model. Hidden actions model the propagation of writes and other operations that modify
shared memory. We use the $\thract$ function to record the operation that each hidden action propagates.
Therefore, for each $f\in Hidden$, we require that $\thract_h(f)\notin Hidden$. ($f$ is short for {\em flush}.)
For example, in TSO, the hidden actions are the flushes, and $\thract_h$ associates with each flush the write that
created the buffer entry which is being flushed.

Flush based memories must satisfy a technical constraint. We require that the effect of a flush be invisible to the thread on whose behalf
the flush is being performed. This captures the idea that flushes are responsible for propagating the effect of
operations from one thread to another, rather than affecting the behaviour of the invoking thread.

\begin{definition}[Local Flush Invisible]
\label{def:flush-invisible}
A memory model $M$ is {\em local flush invisible} if for all histories $h\in M$, actions $a, b, f$ in $h$ 
such that $a = \thract(f)$ and $a\porel_h b\to_h f$, $b$ and $f$ are $M$-independent.
\end{definition}
For the rest of this section, fix a memory model $M$ with thread action function $\thract$. Furthermore, fix an object system $S$,
such that for all $h\in S$, $h\seqslice Mem\in M$. Thus, $S$ is an object system that may contain both external and internal actions.

\begin{definition}[Response Synchronization]
\label{def:response-sync}
Given a history $h$, the {\em response-synchronization relation} of $h$ is
\begin{align}
\rsrel_h = \deprel_h^S \cup \{(f, resp_h(\thract_h(f))) : f\in Hidden\}
\end{align}
\end{definition}

A {\em response-synchronized history} is one where each flush appears before its associated response.
That is, $h\in S$ is response-synchronized if $\rsrel_h\subseteq\to_h$. An object system is {\em response-synchronized
linearizable} (or RS-linearizable) if all its response synchronized histories are linearizable.

It is relatively easy to verify RS-linearizability. The idea is to construct a model of the system such that
response actions are not enabled until the operation's writes have been flushed, and then to prove that the implementation
is linearizable on this stronger model. See \cite{derrick-TSO-lin-2014} for a careful development of the technique.

Operation-race freedom requires that clients provide sufficent synchronization to prevent any thread from observing
that a flush has taken place after its corresponding response action. Definition \ref{def:sync-point} formalizes
which actions count as synchronizing actions, for the purposes of operation-race freedom. 
Operation-race freedom has one key property not shared by
standard notions of data-race freedom: invocations and responses can count as synchronizing actions. This has two advantages.
First, we can reason about the absence of races based on the presence of synchronizing invocations and responses,
rather than being based on low-level memory operations that have synchronization properties. Second, implementations
of concurrent objects are free to employ racey techniques within each operation.

\begin{definition}[Synchronization Point]
\label{def:sync-point}
An action $b$ is a {\em synchronization point} in $h\in S$, if for all actions $a$ such that $a\porel_h b$ or $a=b$,
all actions $c$ such that $thr(c)\neq thr(a)$ and $b\deprel_h^S c$, and all hidden actions $f$ such that $\thract(f)=a$,
not $c\deprel_h^S f$.
\end{definition}
For example, in TSO, barrier operations are synchronization points. This is because such operations
ensure that the issuing thread's write buffer is empty before the barrier is executed. Therefore,
any write before the barrier in program order is flushed before the barrier executes, and so the write's flush
cannot be after the barrier in causal order. For the same reason, locked operations are also synchronization points in TSO.

Under this definition, invocations and responses may also be synchronization points. An invocation is a
synchronization point if its first memory action is a synchronization point, and a
response is a synchronization point if its last memory action is a synchronization point. This is because any external action
is independent of any hidden action.

\begin{definition}[Operation Race]
\label{def:operation-race}
An {\em operation race} (or {\em o-race}) in a history $h$ is a triple $r_0, i, r_1$, where $r_0, r_1$ are responses, $i$ is an
invocation such that $r_0\porel_h i$, $i\deprel_h^S r_1$, $thr(r_0)\neq thr(r_1)$, $obj(r_0)=obj(r_1)$, there is some hidden action
$f$ such that $r_0=resp_h(\thract_h(f))$, and there is no
synchronization point between $r_0$ and $i$ (inclusive) in program order.
\end{definition}
We say that an object system is {\em o-race free} (ORF) if no history has an o-race.

Below we provide an example of an execution containing an o-race. This example and the next use a datastructure called a
{\em seqlock}, another concurrent object optimised for use on TSO, and adapted from an implementation in the Linux kernel
\cite{triang-race-freedom}.
Seqlock is an object providing {\em read} and {\em write} operations with the usual semantics,
except that several values can be read or written in one operation. Seqlock has the restriction that there may
only be one active write operation at a time, but there may be any number of concurrent read operations and reads may
execute concurrently with a write. Seqlock does not use any locking mechanism internally, instead relying on a counter
to ensure that read operations observe a consistent set of values. Seqlock does not use any locked or barrier operations,
and the read operation never writes to any location in memory. Other details of the algorithm do not matter for our purposes.
See \cite{tso-lib-correct-2012} for a complete description.


Consider the behaviour presented in
Figure \ref{fig:orace}, adapted from \cite{triang-race-freedom}. Here, three threads interact using an instance \verb|L|
of the spinlock object, and an instance \verb|S| of seqlock. In this example execution, the flush correponding to the write of
$t_2$'s release operation is delayed until the end of the execution, but the flushes associated with the writes of
$t_1$'s seqlock write operation occur immediately (note that because the seqlock does not use any barrier or locked
operations, this flush could have occurred at any point after the write to memory).
This history is not sequentially consistent.
If it were sequentially consistent, thread $t_2$'s release would need to take effect before thread $t_1$'s write,
which in turn would take effect before thread $t_3$'s read. However, this is inconsistent with the fact that
$t_3$'s try-acquire apears to take effect before thread $t_2$'s release. Because it is not sequentially consistent,
this execution would be impossible if the spinlock and seqlock were both linearizable objects. Therefore, the composition
of spinlock and seqlock do not observationally refine a composition of linearizable objects, for any client capable of
producing this behaviour.
\begin{figure}[t]
{\codesize\tt
\begin{tabular}{l|l|l}
\multicolumn{1}{c}{T1} & \multicolumn{1}{c}{T2} & \multicolumn{1}{c}{T3}\\
\hline
& L.try\_acquire(), & \\
&\ \ \ \ returns true & \\
& L.release() &\\
& S.read(), returns (0,0) \\
S.write(1, 1) & & \\
& & S.read(), returns (1,1) \\
& & L.try\_acquire(), \\
& & \ \ \ \ returns false
\end{tabular}
}
\caption{A racey execution of a spinlock {\tt L} and a seqlock {\tt S}. Operation-race freedom prohibits the
race between $t_2$'s release and read, and $t_3$'s try-acquire.}
\label{fig:orace}
\end{figure}
There is a race between the response of $t_2$'s release operation, the invocation of $t_2$'s subsequent read, and the release
of $t_3$'s try-acquire.

Theorem \ref{theorem:composition} below shows that an ORF multi-object system composed of response-synchronized
linearizable objects is causally linearizable, under one technical assumption.
We require that the objects themselves must not {\em interfere}. That is, each action of each object must be independent of
all potentially adjacent actions of other objects. This constraint is implicit in the standard composition result for
linearizability, and is satisfied by any multi-object system where each object uses regions of shared-memory disjoint from all
the other objects. If a multi-object system does not satsify this property, then one object can affect the behaviour of another
object by modifying its representation. Therefore, a composition of individually linearizable objects may not be linearizable itself.
\begin{definition}[Noninterfering Object System]
\label{def:noninterference}
An object system $S$ is {\em noninterfering} if for all histories $h\in S$, and actions $a, b$ adjacent in $h$, if
$thr(a)\neq thr(b)$ and $obj(a)\neq obj(b)$ then $a$ and $b$ are independent.
\end{definition}

The following lemma shows that the response-synchronization relation is acyclic for an operation-race free object system.
This allows us to prove Theorem \ref{theorem:composition} by applying Lemma \ref{lem:causal-reordering}.
\begin{lemma}[Acyclicity of the Response-synchronization Relation]
\label{lem:resp-sync-acyclic}
If $M$ is a memory model and $C[M]$ is a noninterfering, ORF object system, then for all $h\in C[M]$, the response-synchronisation
relation is acyclic.
\end{lemma}

We can now state our compositionality result.
\begin{theorem}[Composition]
\label{theorem:composition}
Let $X$ be a set of objects and for each $x\in X$, let $T_x$ be a sequential object system. If $M$ is a flush-based memory
and $C[M]$ is an ORF noninterfering multi-object system such that for each $x\in X$, $C[M] \seqslice x$ is response-synchronized
linearizable to $T_x$, then $C[M]$ is causally linearizable to $T = {h : \forall x\in X.~h\seqslice acts(T_x)\in T_x}$,
and thus $C[M]$ observationally refines $C[T]$.
\end{theorem}

\section{Operation-race Freedom on TSO}
\label{sect:tso-causal}

We apply our technique to the well-known {\em total store order} (TSO) memory model, a version of which is implemented by
the ubiqitous x86 processor family. Indeed, we closely follow the formalization of
TSO for x86 given in \cite{owens-x86-tso}. 
We then argue that TSO has the properties required of a flush-based memory, including the local flush invisibility property.
Finally, we demonstrate how to determine whether a client is operation-race free.

We model TSO as a labelled transition system (LTS) $T = \langle S_T, A_T, I_T, R_T\rangle$.
Each state $s\in S_T$ has the form $\langle M, B\rangle$ where
\begin{itemize}
\item $M$ is the contents of shared memory, $M : Loc \to \num$, where $Loc$ is the set of {\em locations}.
\item $B$ records for each thread the contents of its buffer, which is a sequence of location/value pairs.
Thus, $B : T \to (Loc \times \num)^*$.
\end{itemize}
The initial state predicate $I_T$ says only that every buffer is empty
(formally, $\forall t \in T.~B(s) = \sq{}$).
The transition relation $R_T$ is given in Figure \ref{fig:tso-transitions}.
The labels (or {\em actions}) in the set $A_T$ are as follows. For each thread $t\in T$, location $x\in Loc$ and value $v, r\in \num$,
there is a write action $write_t(x, v)$, a read action $read_t(x, r)$, a flush action $flush(t, x, v)$
and a barrier action $barrier_t$. Further, there is a locked action $lock_t(f, x, v, r)$, for each
$f: \num\times\num\to\num$
taken from an appropriate list of {\em read-modify-write} (RMW) operations. Locked actions model the atomic application of an
RMW operation to shared memory. For example, $lock_t(+, x, 1, r)$ models the atomic increment of the value
at $x$, and $r$ is the value in location $x$ immediately before the increment.
The x86 instruction set supports a range of other RMW operations, such as {\em add} and {\em test-and-set}.

\newcommand{\tsostep}[3]{#1\stackarrow{#2}#3}
\newcommand{\rulevspace}{\vspace{1.2em}}

\begin{figure}[t]
{\small
\center
\infer[Read]{\tsostep{(M, B)}{read_t(x, r)}{(M,B)}} {(last\_write(B(t), x) = \bot \wedge M(x) = r) \vee last\_write(B(t), x) = r}
\rulevspace

\infer[Write]{\tsostep{(M, B)}{write_t(x, v)}{(M,\fnup{B}{t}{b'})}} {b' = B(t)\sq{(l, v)}}
\rulevspace

\infer[Flush]{\tsostep{(M, B)}{flush(t, x, v)}{(\fnup{M}{l}{v},\fnup{B}{t}{b'})}} {B(t) = \sq{(l, v)} b'}
\rulevspace

\infer[Barrier]{\tsostep{(M, B)}{barrier_t}{(M,B)}} {B(t)=\sq{}}
\rulevspace

\infer[Locked\hyph RMW]{\tsostep{(M, B)}{lock_t(f, x, v, r)}{(\fnup{M}{x}{f(c, v)}, B)}}
{B(t)=\sq{} & r = M(x)}
}
\caption{\label{fig:tso-transitions} Transition relation of the TSO memory model. If $b$ is a write buffer,
$latest\_write(b, x)$ returns the value of the last write to $x$ in $b$, if it exists, or $\bot$ otherwise.}
\end{figure}
The original formulation of the TSO memory model \cite{owens-x86-tso} has a more complicated representation of locked
actions, arising from the need to model both atomic and nonatomic RMW operations uniformly. We only need to model
atomic RMW operations, so we avoid the additional complexity.

The set of traces of this TSO LTS is prefix-closed and thus forms an object system, which we denote by $TSO$.
The system actions of $TSO$ are just the flush actions, so the $TSO$ $thr$ function returns $\bot$ for flush
actions, and the thread index of all other actions.
The $\thract_h$ function associates with each flush $f$
the write that is being flushed.
$TSO$ has the flush invisibility property, of Definition \ref{def:flush-invisible}, because a flush is independent of any
action of the issuing thread, except for the write that is being flushed (as proved in the full version of the paper).



\begin{wrapfigure}{L}{0.5\textwidth}
{\codesize
\begin{verbatim}
val ensure_init() {
1.  (v0, v1) = X.read();	
2.  if (v0 == null) {
3.    L.acquire();
4.    (v0, v1) = X.read();
5.    if (v0 == null) {
6.      (v0, v1) = initial_value;
7.      X.write(v0, v1);
8.      Barrier();
9.    }
10.   L.release();
11. }
12. return (v0, v1);
}
\end{verbatim}
}
\caption{Pseudocode for a client executing a double checked locking protocol.}
\label{fig:dcl-client}
\end{wrapfigure}

We now explain by example how to check that a client is ORF. Our example is the double-checked locking implementation
presented in Figure \ref{fig:dcl-client}. Double-checked locking is a pattern for lazily initializing a shared object at most once
in any execution. The \verb|ensure_init| procedure implements this pattern. Here, the shared object is represented using a seqlock \verb|X|.
The \verb|ensure_init| procedure first reads the values in \verb|X|, and completes immediately if \verb|X| has already been initialised.
Otherwise, \verb|ensure_init| acquires a spinlock \verb|L| and then checks again whether \verb|X| has already been initialised
(by some concurrent thread), again completing if the initialisation has alread occurred.
Otherwise, \verb|ensure_init| initialises the object, executes a barrier, releases the lock and returns.

To show that this code is ORF, we must
employ knowledge about which invocations and responses of our objects are synchronization points, and which
operations do not execute write actions.
As we described in the discussion after Definition \ref{def:sync-point}, in TSO all barriers and locked operations
are synchronization points. Furthermore, because the try-acquire's only memory operation is a locked operation,
both the invocation and response of try-acquire are synchronization points. Finally, the read operation of seqlock
can never execute a write action.

To show that \verb|ensure_init| has no o-races, we must consider the relationship between each operation, and the next operation in program
order. For each case, we must show that no o-race is possible.
\begin{itemize}
\item The read on Line 1 never executes a write, so its response cannot form an o-race with the subsequent invocation.
\item The response of the acquire on Line 3 is a synchronization point, so it cannot form an o-race with the subsequent read.
\item As with the read on Line 1, the read on Line 4 never executes a write operation, and so its response cannot form an o-race.
\item The write on Line 7 is followed by the barrier on Line 8, so this cannot form an o-race.
\end{itemize}
Note that during this argument, we only need to consider
whether or not the invocation or response of each operation is a synchronization point, or whether the operation never executes
write actions. We do not require any further information about the operation's implementation. Again, this means that operations may
themselves be racey.

\section{Concluding Remarks}
\label{sec:conclusions}

Although the details of the paper are fairly technical the essence of the contribution is simple: how can we use non-linearizable algorithms safely.
The context that we work in here is that of weak memory models, where TSO provides an important example. This work should also be applicable to other flush-based memory models. Such an extension is work for the future.

To enable our multi-object systems to be composed safely we introduced a notion of operation-race freedom.
However, what about non-operation-race free programs? Our formulation provides no composability guaranteess for a family of objects
where even {\em one} of those objects is not response-synchronized. As indicated in Section \ref{sect:flush-mem},
this is a less severe restriction than other proposals based on some notion of data race freedom (because of its
modularity). However, it seems reasonable to expect that some compositionality result would hold for the subset of
response-synchronized objects. Again this is left as future work.

%

\bibliographystyle{plain}
\bibliography{bib}

\begin{thebibliography}{10}

\bibitem{weak-memory-a-tutorial}
Sarita~V Adve and Kourosh Gharachorloo.
\newblock Shared memory consistency models: a tutorial.
\newblock {\em Computer}, 29(12):66--76, Dec 1996.

\bibitem{Alglave:2009:SPA:1481839.1481842}
Jade Alglave, Anthony Fox, Samin Ishtiaq, Magnus~O Myreen, Susmit Sarkar, Peter
  Sewell, and Francesco~Zappa Nardelli.
\newblock {The Semantics of Power and ARM Multiprocessor Machine Code}.
\newblock In L.~Petersen and M.M.T. Chakravarty, editors, {\em DAMP '09}, pages
  13--24. ACM, 2008.

\bibitem{tso-lib-correct-2012}
Sebastian Burckhardt, Alexey Gotsman, Madanlal Musuvathi, and Hongseok Yang.
\newblock Concurrent library correctness on the {TSO} memory model.
\newblock In Helmut Seidl, editor, {\em Programming Languages and Systems},
  volume 7211 of {\em Lecture Notes in Computer Science}, pages 87--107.
  Springer Berlin Heidelberg, 2012.

\bibitem{mech-verif-proof-2011}
John Derrick, Gerhard Schellhorn, and Heike Wehrheim.
\newblock Mechanically verified proof obligations for linearizability.
\newblock {\em ACM Trans. Program. Lang. Syst.}, 33(1):4:1--4:43, January 2011.

\bibitem{derrick-TSO-lin-2014}
John Derrick, Graeme Smith, and Brijesh Dongol.
\newblock {Verifying Linearizability on TSO Architectures}.
\newblock In Elvira Albert and Emil Sekerinski, editors, {\em Integrated Formal
  Methods}, volume 8739 of {\em Lecture Notes in Computer Science}, pages
  341--356. Springer International Publishing, 2014.

\bibitem{abstract-for-conc-obj-2010}
Ivana Filipović, Peter O’Hearn, Noam Rinetzky, and Hongseok Yang.
\newblock Abstraction for concurrent objects.
\newblock {\em Theoretical Computer Science}, 411(51–52):4379 -- 4398, 2010.
\newblock European Symposium on Programming 2009ESOP 2009.

\bibitem{gotsman2012show}
Alexey Gotsman, Madanlal Musuvathi, and Hongseok Yang.
\newblock Show no weakness: Sequentially consistent specifications of tso
  libraries.
\newblock In {\em Distributed Computing}, pages 31--45. Springer, 2012.

\bibitem{linearizability-toplas}
Maurice~P Herlihy and Jeannette~M Wing.
\newblock Linearizability: A correctness condition for concurrent objects.
\newblock {\em ACM Trans. Program. Lang. Syst.}, 12(3):463--492, July 1990.

\bibitem{triang-race-freedom}
Scott Owens.
\newblock {Reasoning about the Implementation of Concurrency Abstractions on
  x86-TSO}.
\newblock In Theo D’Hondt, editor, {\em ECOOP 2010 – Object-Oriented
  Programming}, volume 6183 of {\em Lecture Notes in Computer Science}, pages
  478--503. Springer Berlin Heidelberg, 2010.

\bibitem{owens-x86-tso}
Scott Owens, Susmit Sarkar, and Peter Sewell.
\newblock {A Better x86 Memory Model: x86-TSO}.
\newblock In Stefan Berghofer, Tobias Nipkow, Christian Urban, and Makarius
  Wenzel, editors, {\em Theorem Proving in Higher Order Logics}, volume 5674 of
  {\em Lecture Notes in Computer Science}, pages 391--407. Springer Berlin
  Heidelberg, 2009.

\end{thebibliography}

\newpage
\appendix

\section{Appendix}


\begin{lemma}
\label{Alem:causal-ordering-props}
For all $h\in S$, and causally equivalent histories $h'$,
\begin{enumerate}
\item $h$ is thread equivalent to $h'$,
\item $h'\in S$,
\item $a \deprel_h b \iff a \deprel_{h'} b$.
\end{enumerate}
\end{lemma}
\Proof The proof of the first two properties is a simple induction on the number of transpositions required to get from $h$ to $h'$.
For the third property, observe that causal equivalence is an equivalence, so the set of histories causally equivalent
to $h$ and that causally equivalent to $h'$ is the same. Hence, the order derived form each set is the same.
\qed

\begin{lemma}
\label{Alem:causal-reordering}
Let $h$ be a history, and let $<$ be a strict partial order on the events of $h$ such that $\deprel_h\subseteq <$. Then 
there exists an $h'$ causally equivalent to $h$ such that for all events $a,b$ in $h$ (equivalently in $h'$)
$a < b$ implies $a\to_{h'} b$.
\end{lemma}
\Proof A {\em mis-ordered pair} is a pair of actions $a, b$ such that $a < b$ but $b\to_h a$. Note that for any mis-orderd pair
$a, b$, because $\deprel_h\subseteq <$, we have not $b\deprel_h a$.
Let $a, b$ be the mis-ordered pair with the least difference in indexes. 
If $a$ and $b$ are adjacent, then they must be independent, so we can simply reorder them, strictly
reducing the number of mis-ordered pairs.

If $a$ and $b$ are not adjacent, we reduce the size of the gap between them, without creating new mis-ordered pairs. Eventually,
the pair will be adjacent, and we can then reduce the number of mis-ordered pairs. Let $g$ be the (nonempty) sequence
between $b$ and $a$. For any action $c$ in $g$ let $g_c$ be the prefix of $\sq{b}\seqcomp g$ that ends just before $c$, and let
$g'_c$ be the suffix of $g\seqcomp\sq{a}$ beginning just after $c$. 
If there is any action $c$ in $g$ such that for all $d$ in $g_c$ not $d < c$, then reorder $c$ before $b$, reducing the gap
between $b$ and $a$. Clearly, this does not create new mis-ordered pairs, and the resulting history is causally
equivalent to $h$ because $\deprel_h\subseteq <$.
Likewise, if there is any $c$ in $g$ such that for all $d'$ in $g'_c$ not $c < d$, then reorder $c$ after $a$.

In the remaining case we have for all $c$ in $g$, there is some $d$ in $g_c$ such that $d < c$, and there is some $d'$ in
$g'_d$ such that $c < d$. But then there is a chain $b < d_0 < \ldots < d_m < c$ and a chain $c < d'_0 < \ldots < d'_n < a$
(where each $d_i$ is in $g_c$ and each $d'_i$ is in $g'_c$), and thus $b < a$, contradicting the irreflexivity of $<$.
\qed

\subsection{Tagged Events}

An {\em event index} is a pair $(t, i)$ where $t\in T$ and $i$ is
a natural number. Given a history $h$, we create a corresponding {\em event sequence} by tagging each action 
in $h$ with an event index. Each action $a$ such that $thr(a)\in T$ is tagged with $(thr(a), i)$ where $i$ is the index
of the action within $h\seqslice thr(a)$. Each action $a$ such that $thr(a)=\bot$ is tagged with $(t, i)$ where
$t=thr(\thract_h(a))$ and $i$ is the index of $\thract_h(a)$ in $h\seqslice t$. (Note that a thread event may share its tag with
a hidden event, but then the associated action will be different, and thus uniqueness is preserved.)

\subsection{Properties of Clients}

\begin{lemma}
\label{Alem:client-monotonic}
For all object systems $O$, clients $C, C'$ such that $C'\subseteq C$, $C'[O]\subseteq C[O]$
\end{lemma}
Immediate from definitions.
\qed

\begin{lemma}
\label{Alem:causal-decomp}
For all object systems $M$, clients $C$, and memory actions (indexed by distinct threads) $a$, $b$,
histories $g=g_0\seqcomp \sq{a, b}\seqcomp g_1\in C[O]$ and $g'=g_0\seqcomp \sq{b, a}\seqcomp g_1\notin C[O]$
then $g\seqslice acts(O)\in O$ and $g'\seqslice acts(O)\notin O$.
\end{lemma}
$g_0\seqslice C[O] \implies g_0\seqslice acts(O) \in O$, by the defintion of $C[O]$. To see that $g'\seqslice acts(O) \notin O$, assume
otherwise and observe that because for all threads $t$, $g'\seqslice t=g\seqslice t \in C$, and so $g'$ would be in $C[O]$, contrary
to assumptions.
\qed

\begin{lemma}\label{Alem:independence-preserved}
For all object systems $O$, clients $C, C'$ such that $C'\subseteq C$, and for all actions
$a, b$ with distinct thread identifiers, if $a$ and $b$ are independent in $C[O]$ then they are
independent in $C'[O]$.
\end{lemma}
Assume that $a$ and $b$ are independent in $C[O]$ but not in $C'[O]$. First, note that because $a$
and $b$ are independent in $C[O]$, they must have diferent thread identifiers.
Second, observe that there are histories $g, g', g_0, g_1$ such that $g=g_0\seqcomp \sq{a, b}\seqcomp g_1\in C'[O]$
but $g'=g_0\seqcomp \sq{b, a} \seqcomp g_1\notin C'[O]$.
But then, by Lemma \ref{Alem:causal-decomp},
\[
\label{Aprop:g-prime-illegal}
g'\seqslice acts(O) \notin O
\]
Because $C'[O]\subseteq C[O]$, it must be that $g\in C[O]$, and because $a$ and $b$ are independent in
$C[O]$ we have $g'\in C[O]$, which contradicts \ref{Aprop:g-prime-illegal}.
\qed

\begin{lemma}
\label{Alem:client-causal-pres}
For all object systems $O$, clients $C$, $C'$ where $C' \subseteq C$,
and for all $h \in C'[O]$, $\deprel{h}^{C'[O]} \subseteq \deprel_h^{C[O]}$.
Thus, if two actions $a$ and $b$ can be reordered in an execution of the more general client,
then they can be reordered in the execution when regarded as a history of the less general client.
\end{lemma}
Lemma \ref{Alem:independence-preserved} shows that
independence in $C[O]$ is included in independence in $C'[O]$. Therefore, causal equivalence in $C'[O]$ is
coarser than that of $C[O]$. Hence, $\deprel{h}^{C'[O]} \subseteq \deprel_h^{C[O]}$.
\qed

\begin{lemma}
Let $C$ be a client, $S$ be an object system, and $T$ be a sequential object system such that $C[S]$ is causally
linearizable to $T$. Then for all clients $C'\subseteq C$, $C'[S]$ is causally linearizable to $T$.
\end{lemma}
By Lemma \ref{Alem:client-causal-pres}, causal equivalence of $C'[S]$ is coarser than the causal equivalence of $C[S]$. 
\qed

\begin{theorem}[Causal Linearizability Implies Observational Refinement]
\label{Atheorem:causal-lin-sound}
Let $T$ be a sequential object system, and let $T'$ be its set of linearizable histories. Let $S$ be an object system
such that $acts(T) \cap Ext = acts(S) \cap Ext$. If $C[S]$ is causally linearizable to $T$, then $S$ observationally refines
$T'$ for $C$.
\end{theorem}
We must show that for each $h \in C[S]$, there is some history $h' \in C[T']$ such that $h\seqslice Ext$ and $h'\seqslice Ext$
are thread equivalent.
Because $C[S]$ is causally linearizable to $T$, for each $h\in C[S]$, there is some $C[S]$-causally-equivalent history
$h'$ such that $h' \seqslice Ext$ is linearizable to $T$, and therefore $h'\seqslice Ext$ is in
$T'$. Note that because $h$ and $h'$ are causally equivalent, they are both thread equivalent. This implies that
$h'\seqslice t\in C$ for every $\in T$. Thus $h' \in C[T']$. This completes the proof.
\qed

\subsection{Flush-based Memory and Operation-race Freedom}

\begin{lemma}[Acyclicity of the Response-synchronization Relation]
\label{Alem:resp-sync-acyclic}
If $M$ is a memory model and $C[M]$ is a noninterfering, ORF object system, then for all $h\in C[M]$, the response-synchronisation
relation is acyclic.
\end{lemma}
Given a cycle in the response-synchronization relation, we construct an operation race. Assume that there is a cycle in the
response synchronization relation. That is, assume there exists
a history $h$, a thread $t$, a response $r$ with $t=thr(r)$, and a system action $f$ such that $r=resp_h(\thract_h(f))$ but
$r\deprel_h f$, and therefore $h$ has a cycle in the response-synchronization relation. Because $f$ is independent of
all $t's$ actions occurring after $r$ in program order, there must be some other thread $t'\neq t$
and action $b$ with $t'=thr(b)$, $r\deprel_h b$ (1) and $b$ and $f$ are not independent and $b\to_h f$
(and thus $b\deprel_h f$) (2). Because the response of
one thread is independent of all actions of all distinct threads (since $M$ is a memory model), (1) implies
that there must be some memory action $a$ such that $r\porel_h a$ and $a\deprel_h b$.

We show that $r, inv_h(a)$ and $resp_h(b)$ form an operation race. If there is some synchronization point between
$r$ and $a$ (inclusive) in program order, then because $\thract(f)\porel_h r$, we would have not $b\deprel_h f$.
So there is no such synchronization point. Furthermore, because $b\porel_h resp_h(b)$ we have $b\deprel_h resp_h(b)$
and therefore $inv_h(a)\porel_h a\deprel_h b\porel_h resp_h(b)$, we have $inv_h(a)\deprel_h resp_h(b)$.
By construction, we have $r=resp_h(\thract_h(f))$.
Finally, $obj(resp_h(b)) = obj(r)$, because $C[M]$ is a noninterfering object system and $b$ and $f$ are not
independent.
\qed

\begin{theorem}[Composition]
\label{Atheorem:composition}
Let $X$ be a set of objects and for each $x\in X$, let $T_x$ be a sequential object system. If $M$ is a flush-based memory
and $C[M]$ is an ORF noninterfering multi-object system such that for each $x\in X$, $C[M] \seqslice x$ is response-synchronized
linearizable to $T_x$, then $C[M]$ is causally linearizable to $T = {h : \forall x\in X.~h\seqslice acts(T_x)\in T_x}$,
and thus $C[M]$ observationally refines $C[T]$.
\end{theorem}
\Proof Lemma \ref{Alem:resp-sync-acyclic} establishes that the ORF freedom
of $C[M]$ implies that the response-synchronization relation is acyclic, and therefore its transitive closure is
a strict partial order. Thus, by Lemma \ref{Alem:causal-reordering} for each
$h\in C[M]$ there is some
$h'\in C[M]$ $C[M]$-causally equivalent to $h$, such that $\rsrel_{h'} \subseteq\to_{h'}$. Now, for every $x\in X$,
$h'\seqslice x$ is response-synchronized and therefore $h'\seqslice acts(T_x)\cap Ext$ is linearizable to $T_x$. Now,
the classic argument for the locality of linearizability from \cite{linearizability-toplas} can be used to show
that $h'\seqslice acts(T)\cap Ext$ is linearizable to $T$.
\qed

\begin{lemma}[Flush Invisibility for TSO]
\label{Alem:tso-flush-invis}
$TSO$ has the flush invisibility property, of Definition \ref{def:flush-invisible}.
\end{lemma}
\Proof Each flush $f$ is independent of all reads and writes of the same process, except for the write being flushed.
$f$ is not independent of barriers or locked operations, but if $a\porel_h b\to_f$, then $b$ cannot be a barrier or locked operation.
\qed

\begin{lemma}[Synchronization Points in TSO]
\label{lem:tso-abs-barriers}
In TSO, all barriers and locked operations are synchronization points.
\end{lemma}
\Proof 
In TSO, if $b$ is a barrier or locked operation and $a\porel b$, then for any flush $f$ such that $\thract_h(f)=a$,
$f\to_h b$, so not $b\deprel_h f$.
\qed

\begin{lemma}[Synchronization Points of Spinlock]
\label{lem:lock-sync-points}
Both the invocation and response
of any acquire operation is a synchronization points on TSO. The invocation of any release operation is a synchronization point on TSO
so long as it is not followed in program order by a read action before any synchronising action.
\end{lemma}
The first and last operation of acquire is a locked operation, so the result is straightforward for the acquire operation.

Let $h\in TSO$ be a history and $i$ be an invocation of release in $h$. Let $c$ be an action of another thread such that
$i\deprel_h c$. Because $i$ is an invocation it is independent of the actions of any other thread, so we must have
$w_r\deprel_h c$, where $w_r$ is the write of the release operation. Let $w$ be any write such that $w\porel_h i$. but then the flush of
$w$ appears before that of $w_r$ in realtime order.
\qed

\end{document}